\documentclass[twocolumn,aps,superscriptaddress,showpacs]{revtex4}
\usepackage{amssymb}
\usepackage{amsmath}
\usepackage{mathrsfs}
\usepackage{graphicx}
\usepackage{subfigure}
\usepackage[normalem]{ulem}
\usepackage[dvips]{color}
\usepackage{xcolor}
\usepackage{hyperref}

\hypersetup{hypertex=true,
colorlinks=true,
linkcolor=magenta,
citecolor=magenta,
urlcolor=cyan} 

\begin{document}

\title{Axion effects on hybrid stars: earlier hadron-quark phase transition and larger quark-matter cores}
\author{Yu-Heng Liu} 
\affiliation{Science School, Qingdao University of Technology, Qingdao 266520, China}
\affiliation{The Research Center of Theoretical Physics, Qingdao University of Technology, Qingdao 266033, China}
\author{He Liu}
\email{liuhe@qut.edu.cn}
\affiliation{Science School, Qingdao University of Technology, Qingdao 266520, China}
\affiliation{The Research Center of Theoretical Physics, Qingdao University of Technology, Qingdao 266033, China}
\author{Hong-Ming Liu}
\email{liuhongming13@126.com}
\affiliation{Science School, Qingdao University of Technology, Qingdao 266520, China}
\affiliation{The Research Center of Theoretical Physics, Qingdao University of Technology, Qingdao 266033, China}
\author{Min Ju}
\email{jumin@upc.edu.cn}
\affiliation{School of Science, China University of Petroleum (East China), Qingdao 266580, China}
\author{Xu-Hao Wu}
\email{wuhaoysu@ysu.edu.cn}
\affiliation{School of Science, Yanshan University, Qinhuangdao 066004, China}
\author{Peng-Cheng Chu}
\email{kyois@126.com}
\affiliation{Science School, Qingdao University of Technology, Qingdao 266520, China}
\affiliation{The Research Center of Theoretical Physics, Qingdao University of Technology, Qingdao 266033, China}
\date{\today}

\begin{abstract}
We investigate the effects of axions on the properties of hybrid stars and the possible hadron-quark phase transition in their cores within the Maxwell construction. The hadronic phase is described by the improved isospin- and momentum-dependent interaction (ImMDI) model, while the quark phase is modeled by a three-flavor Nambu--Jona-Lasinio (NJL) model with an axion-modified Kobayashi-Maskawa-'t Hooft (KMT) interaction. Based on the obtained equation of state, we analyze the properties of hybrid-star matter, including the speed of sound and the polytropic index, as well as the structural characteristics of hybrid stars, including the mass-radius relation and the properties of the quark core. Our study shows that the axion field $a/f_a$ slightly stiffens the equation of state of quark matter but significantly shifts the hadron-quark phase transition to lower baryon densities. These two effects compete in influencing the equation of state of hybrid-star matter, and may even soften it, which is different from the effect of the vector coupling interaction. Consequently, the axion field has only a weak influence on the maximum mass of hybrid stars, but significantly increases the mass and radius of the quark core, especially for weak vector coupling. These results indicate that, within the present Maxwell construction, the primary role of the axion is to promote an earlier hadron-quark transition at low densities and to enlarge the quark core.
\end{abstract}
\maketitle

\section{Introduction}
The equation of state (EOS) of strongly interacting matter is among the central topics in nuclear physics and astrophysics, since it encodes the thermodynamic properties of dense nuclear matter and
determines the structure of compact objects~\cite{Web05}. The behavior of cold and dense matter at supranuclear densities is crucial for understanding the phase diagram of quantum chromodynamics (QCD), where the phase transition from hadronic to quark matter may take place~\cite{Aok06}. In recent years, despite sustained progress in theoretical methods and experimental studies of nuclear physics, the EOS of strongly interacting matter remains far from settled~\cite{Xie21,Han23}. A major source of uncertainty is that first-principles lattice QCD calculations are not presently applicable in the cold and dense regime because of the sign problem~\cite{Nag22}. However, some effective theoretical approaches can yield predictions of a different nature for the hadron-quark phase transition, ranging from a smooth crossover to a strong first-order transition, or even an extended mixed phase~\cite{Kum24,Fuk25}. Therefore, the structural characteristics of neutron stars composed of cold and dense matter, as well as whether a hadron-quark phase transition exists in their cores, remain an open question~\cite{Alf19}.

In recent years, multi-messenger astronomical observations have provided increasingly strict constraints on the equation of state of dense matter. Precision mass measurements of heavy pulsars have established that the EOS must be sufficiently stiff to support neutron stars (NSs) with masses of about two solar masses. For example, the measurements of PSR J1614-2230 and PSR J0348+0432 have precisely determined their masses to be $1.908 \pm 0.016 M_{\odot}$~\cite{Dem10,Fon16,Arz18} and $2.01\pm 0.04 M_{\odot}$~\cite{Ant13}, respectively. Pulse-profile analyses from Neutron star Interior Composition Explorer (NICER) have also provided precise mass and radius constraints for several pulsars, including PSR J0030+0451~\cite{Ril19,Mil19} and PSR J0740+6620~\cite{Cro20,Ril21,Mil21}. In addition, the binary neutron-star merger GW170817 provided the first direct tidal deformability constraint on dense matter, significantly limiting the allowed stiffness of the EOS in the intermediate-density region and disfavoring models that are either too soft or too stiff~\cite{Abb17}. These multi-messenger constraints can be summarized as follows: the EOS must support massive compact stars, remain compatible with radii of about 11-13 km, and reproduce the tidal deformability bounds inferred from gravitational-wave data~\cite{Ann20}. These combined constraints have greatly highlighted the necessity of establishing realistic models of dense matter, particularly to better understand whether strong first-order phase transitions can occur in compact stars and how such phase transitions manifest themselves in observable stellar properties~\cite{Bau19}.

Meanwhile, some studies have suggested that physics beyond the Standard Model may also leave observable imprints on dense matter~\cite{Sed16,Sed19,Hoo18}. Among the compelling beyond-Standard-Model particles is the QCD axion, originally introduced through the Peccei-Quinn ($PQ$) mechanism~\cite{PQ77a,PQ77b} to resolve the strong charge-parity ($CP$) problem and now widely regarded as a promising dark-matter candidate. Recent work suggests that the axion is associated with stellar evolution~\cite{Raf91,Jan96} and the anomalous cooling problem of stars~\cite{Sed16,Sed19}, and may be detectable through binary neutron-star merger events~\cite{Dev24,Dia24}. Consequently, this exotic particle could also be produced inside neutron stars and affect their properties. Because axions couple to the QCD sector, they can modify the thermodynamics of dense quark matter~\cite{Chu24,Gon24,Liu25,Lu25} and thus alter the phase structure of compact stars. This makes axions particularly interesting for current studies of their effects on the properties of compact stars and the possible hadron-quark phase transition in the cores of these stars~\cite{Lop22,Liu25}.

The hadron-quark phase transition is predicted to exist in the core of compact stars based on various theoretical approaches. In the Maxwell construction, the hadronic phase transitions to the quark phase through a first-order transition, with the two phases sharing the same pressure and baryon chemical potential but differing in the electron chemical potential. The Gibbs construction is more commonly used to describe a mixed phase of hadrons and quarks, allowing the two coexisting phases to be charged separately~\cite{Gle92}. Both the Maxwell and Gibbs constructions involve only bulk contributions, neglecting finite-size effects such as surface and Coulomb contributions. The possible geometric structures of the mixed phase have been extensively discussed in Refs.~\cite{Mar07,Na12,Wu19,Lug21,Ju21}. However, the large uncertainties in the structure and density range of the mixed phase are still present. In the present work, we adopt the Maxwell construction to describe the hadron-quark phase transition. Compared to the Gibbs construction, the Maxwell approach describes a sharp first-order transition~\cite{Con23} and is therefore well suited for examining how axions affect the first-order transition, the stiffness change of the equation of state, and the resulting star structure~\cite{Sen24}.

Specifically, we use different theoretical models to describe nuclear matter and quark matter, respectively, and construct the equation of state for hybrid-star matter through the Maxwell construction. Based on the obtained equation of state, we systematically analyze the properties of hybrid-star matter, such as the speed of sound and the polytropic index, as well as the structural characteristics of hybrid stars, including the mass-radius relation and the properties of the quark core. In the model describing quark matter, we introduce the effect of the axion field and focus on its influence on the first-order hadron-quark phase transition and on the quark core. Our study is complementary to recent work on hybrid stars with large quark cores as well as to explorations of the role of axions in extreme astrophysical environments.

\section{Theoretical model and methods}
For the quark phase, we employ the three-flavor Nambu--Jona-Lasinio (NJL) model as an effective low-energy description of QCD~\cite{NJL61a,NJL61b}. This model captures the essential chiral dynamics of quark matter and has been widely applied to studies of dense QCD matter~\cite{Chr25,Pin23,Alb24}. It also provides a convenient framework for incorporating axion-induced $CP$-violating effects through the Kobayashi-Maskawa-'t Hooft (KMT) determinant interaction~\cite{Boe08,Cha12,Abh21,Liu25}. The Lagrangian density of the three-flavor NJL model in the presence of the axion field is then given by
\begin{equation}
\begin{aligned}
\mathcal{L}&=\ \bar{\psi}(i\!\not\!\partial-\hat m)\psi +\frac{G_S}{2}\sum_{a=0}^{8}\left[(\bar\psi\lambda_a\psi)^2+(\bar\psi i\gamma_5\lambda_a\psi)^2\right] \\
&-K\left\{e^{i\frac{a}{f_a}}\det[\bar\psi(1+\gamma_5)\psi]+e^{-i\frac{a}{f_a}}\det[\bar\psi(1-\gamma_5)\psi]\right\} \\
&-\frac{G_V}{2}\sum_{a=0}^{8}\left[(\bar\psi\gamma^\mu\lambda_a\psi)^2+(\bar\psi\gamma_5\gamma^\mu\lambda_a\psi)^2\right].
\end{aligned}
\end{equation}
Here $\psi=(u,d,s)^T$ denotes the quark field, $\hat m={\rm diag}(m_u,m_d,m_s)$ is the current-quark mass matrix, and $\lambda_a$ are the flavor SU(3) Gell-Mann matrices with $\lambda_0=\sqrt{2/3}\,I$. The couplings $G_S$, $G_V$, and $K$ stand for the scalar, vector, and six-fermion KMT~\cite{tH76} interaction strengths, respectively. The scalar channel governs dynamical chiral symmetry breaking, the vector channel provides repulsion at high density, and the KMT term breaks the axial $U(1)_A$ symmetry and couples different flavors~\cite{Hat94}. The chiral rotation angle $a/f_a$ in the KMT term originates from the axion field contribution and is also known as the axion angle $\theta$, where $f_a$ is the decay constant of the axion and also represents the $PQ$ symmetry-breaking scale. The effects of strong $CP$ violation for $a/f_a \neq 0$ on the phase diagram of strongly interacting matter have been studied in the NJL model~\cite{Cha12,Lu19}. The results indicate that a nonzero $a/f_a$ term leads to the possibility of a uniform $P$-violating quark-antiquark condensate in the pseudoscalar channel. In this case, in the Mean-field approximation, besides the scalar condensates $\sigma_i=\langle \bar\psi_i\psi_i\rangle$, one must also introduce the pseudoscalar condensates~\cite{Cha12}
\begin{equation}
\eta_i=\langle \bar\psi_i i\gamma_5\psi_i\rangle.
\end{equation} 
For $a/f_a\neq 0$, the KMT term mixes the scalar and pseudoscalar channels, so the constituent mass of flavor $i$ is given by both contributions
\begin{equation}
M_i=\sqrt{(M_i^s)^2+(M_i^p)^2},
\end{equation} where $M_i^s$ and $M_i^p$ are the scalar and pseudoscalar mass contributions, respectively~\cite{Abh21}. Both two contributions to constituent quark masses are determined from the gap equations
\begin{equation}
\begin{aligned}
M_i^s&=m_i+2G_S\sigma_i+2K\left[\cos\left(\frac{a}{f_a}\right)(\sigma_j\sigma_k-\eta_j\eta_k)\right.\\
&\left.+\sin\left(\frac{a}{f_a}\right)(\sigma_j\eta_k+\eta_j\sigma_k)\right],
\end{aligned}
\end{equation}
\begin{equation}
\begin{aligned}
M_i^p&=2G_S\eta_i+2K\left[\cos\left(\frac{a}{f_a}\right)(\eta_j\sigma_k+\sigma_j\eta_k) \right.\\
&\left.+\sin\left(\frac{a}{f_a}\right)(\eta_j\eta_k-\sigma_j\sigma_k)\right],
\end{aligned}
\end{equation}where $(i,j,k)$ denotes a cyclic permutation of $(u,d,s)$. These equations show explicitly how the axion field couples the scalar and pseudoscalar sectors through the $\cos(a/f_a)$ and $\sin(a/f_a)$ terms. The corresponding condensates are obtained self-consistently from
\begin{equation}
\begin{aligned}
\sigma_i=-2N_c\Big[\int_0^\Lambda\frac{d^3p}{(2\pi)^3}\frac{M_i^s}{E_i}\!-\!\int\frac{d^3p}{(2\pi)^3}\frac{M_i^s}{E_i}(f_i+\bar f_i)\Big],
\end{aligned}
\end{equation}
\begin{equation}
\begin{aligned}
\eta_i=2N_c\Big[\int_0^\Lambda\frac{d^3p}{(2\pi)^3}\frac{M_i^p}{E_i}\!-\!\int\frac{d^3p}{(2\pi)^3}\frac{M_i^p}{E_i}(f_i+\bar f_i)\Big],
\end{aligned}
\end{equation}
where $f_i$ and $\bar{f_i}$ are respectively the Fermi distribution functions of particle and antiparticle, and $E_i=\sqrt{p^2+M_i^2}$ is the single-particle energy.

The thermodynamic potential density from the finite-temperature field theory can be written as
\begin{equation}
\begin{aligned}[t]
\Omega &= \!-\!2N_c\!\sum_{i}\Big\{\!\int_0^{\Lambda}\!\frac{d^3p}{(2\pi)^3}E_i\!+\!\int\!\frac{d^3p}{(2\pi)^3}\bigl[T\ln(1\!+\!e^{\!-\!\beta(E_i\!-\!\tilde{\mu}_i)})\\
&\!+\!T\ln(1+e^{-\beta(E_i+\tilde{\mu}_i)})\bigr]\!\Big\}\!+\!\sum_i\bigl[G_S(\sigma_i^2+\eta_i^2)-G_V\rho_i^2\bigr]\\
&\!-\!4K\Bigl[ \cos\frac{a}{f_a}\bigl(\sigma_u\sigma_d\sigma_s-\sigma_s\eta_u\eta_d-\sigma_u\eta_d\eta_s-\sigma_d\eta_u\eta_s\bigr)\\
&\!-\!\sin\frac{a}{f_a}\bigl(\eta_u\eta_d\eta_s-\sigma_u\sigma_s\eta_d-\sigma_d\sigma_s\eta_u-\sigma_u\sigma_d\eta_s\bigr)\Bigr],
\end{aligned}
\end{equation}where the prefactor $2N_c=6$ accounts for spin and color degeneracies of quark. The quantity $\beta=1/T$ is the inverse temperature, $\tilde\mu_i=\mu_i-2G_V\rho_i$ is the effective chemical potential, and $\Lambda$ is the three-momentum cutoff introduced to regularize the ultraviolet-divergent vacuum contribution~\cite{Fuk04}. The pressure and energy density can be derived as
\begin{equation}
P_Q=-(\Omega-\Omega_0),  \qquad   \varepsilon_Q=\sum_{i=u,d,s}\mu_i\rho_i-P_Q,
\end{equation}
where $\Omega_0$ is introduced to ensure that both the pressure and energy density vanish in vacuum. In the present calculations, the parameter set of the NJL sector is taken as $m_u=m_d=5.5$ MeV, $m_s=135.7$ MeV, $G_S\Lambda^2=3.67$, $K\Lambda^5=9.29$, $\Lambda=631.4$ MeV~\cite{Bub05,Bra13}, while $G_V$ and $a/f_a$ are treated as control parameters for the stiffness of quark matter and the strength of axion effects, respectively.  

In the quark phase of a hybrid star, in addition to quark matter, leptons may also be present. In our work, the leptons include both electrons and muons with their masses $m_e=0.511$ MeV and $m_{\mu}=106$ MeV, respectively. The energy density and the pressure can be given as
\begin{equation}
\varepsilon_L = \sum_{i=e,\mu} \frac{1}{\pi^2} \int_0^{p_f^i} \sqrt{p^2 + m_i^2} p^2dp,
\end{equation}
\begin{equation}
P_L = \sum_{i=e,\mu} \mu_i \rho_i - \varepsilon_L,
\end{equation}
where $p_f^i = (3\pi^2 \rho_i)^{\frac{1}{3}}$ is the lepton Fermi momentum. The total energy density and pressure of
quark phase including the contributions from both quarks and leptons are given by
\begin{equation}
\varepsilon^Q = \varepsilon_Q + \varepsilon_L,
\end{equation}
\begin{equation}
P^Q = P_Q + P_L. 
\end{equation}
For a system in equilibrium, these particles must satisfy baryon number conservation, charge neutrality, and $\beta$-equilibrium conditions, which can be written as
\begin{equation}
\rho_B = \frac{1}{3}(\rho_u + \rho_d + \rho_s), 
\end{equation}
\begin{equation}
\frac{2}{3}\rho_u - \frac{1}{3}\rho_d - \frac{1}{3}\rho_s - \rho_e - \rho_\mu = 0, 
\end{equation}
\begin{equation}
\mu_s=\mu_d = \mu_u + \mu_e, \quad \mu_\mu = \mu_e.
\end{equation}

Due to the difficulty in simultaneously satisfying the absolute stability condition of strange quark matter (SQM) and obtaining an equation of state sufficiently stiff to support massive compact stars, models of pure quark stars (QSs) are strongly constrained. Thus, hybrid stars with a hadronic layer provide a more realistic framework for investigating axion effects under these stiffness requirements~\cite{Far84,Bay18}. In such a picture, the outer layer of the core is nuclear matter, while quark degrees of freedom appear in the inner core only when the density exceeds the deconfinement threshold. For the hadronic phase, we employ the improved isospin- and momentum-dependent interaction (ImMDI) model~\cite{Xu15}. The ImMDI model provides an effective description of asymmetric nuclear matter with explicit isospin and momentum dependence in the single-particle potential, and it has been widely applied to the study of dense matter in neutron stars as well as to isospin-sensitive observables in intermediate-energy heavy-ion collisions~\cite{Xu19,Liu22}. Its parameters are constructed so as to reproduce the empirical properties of nuclear matter around saturation density while allowing controlled variations of the symmetry energy and the momentum dependence of the isovector potential. In particular, the three parameters $x$, $y$, and $z$ are introduced to tune the slope parameter of the symmetry energy, the momentum dependence of the symmetry potential, and the symmetry energy at saturation density, respectively~\cite{Xu15,Liu22}. In the present work, in order to focus on the axion effects on the quark phase and the hadron-quark phase transition, we adopt a fixed parameter set in the ImMDI model, $x=-0.1$, $y=64$ MeV, $z=-5$~\cite{Liu22,Liu25}. This choice gives a hadronic equation of state that is compatible with the basic empirical constraints on nuclear matter and remains sufficiently stiff in the density region relevant for massive neutron stars. The corresponding properties of nuclear matter at saturation density are taken as $\rho_0=0.16 {\rm fm}^{-3}$, $E_0(\rho_0) = -15.9$ MeV, $K_0=275$ MeV, $E_{\rm sym}(\rho_0) = 32.5$ MeV, $L = 106$ MeV, $m_s^\ast = 0.7 m_N$, $U_{0,\infty} = 75$MeV, where $\rho_0$ is the saturation density, $E_0(\rho_0)$ is the binding energy per nucleon at saturation, $K_0$ is the incompressibility coefficient, $E_{\rm sym}(\rho_0)$ is the symmetry energy at saturation, $L$ is its slope parameter, $m_s^\ast$ is the isoscalar effective mass, and $U_{0,\infty}$ is the single-particle potential in symmetric nuclear matter at infinitely large nucleon momentum. The detailed calculation of the energy density and pressure of the hadronic phase described by the ImMDI model, as well as further discussions, can be found in our previous work~\cite{Liu22}.

\begin{figure}[h]
\centering
\includegraphics[width=0.9\columnwidth]{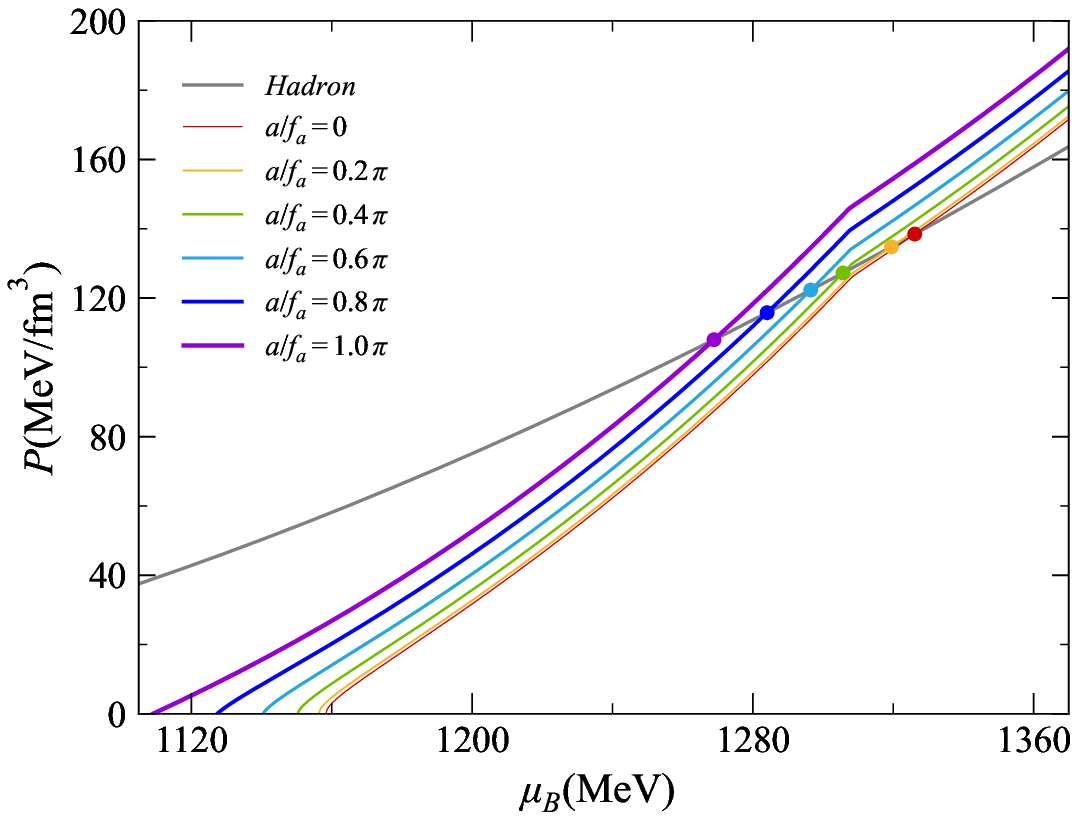}
\caption{Pressure $P$ as a function of baryon chemical potential $\mu_B$ for hadronic matter and quark matter with different scaled axion fields $a/f_a$ at $G_V=0$.}
\label{fig1}
\end{figure}

For the hadron-quark phase transition, the Maxwell construction requires that the pressures and baryon chemical potentials of the two phases be equal. As shown in Fig.~\ref{fig1}, we plot the pressure as a function of baryon chemical potential for the hadronic phase (grey solid line) and for the quark phase (color solid lines) with different scaled axion fields at $G_V=0$. The phase transition point is identified by the intersection between the hadronic curve and the corresponding quark curve. At the intersection, the two phases satisfy the phase-equilibrium condition
\begin{equation}
P^H(\mu_B)=P^Q(\mu_B),
\end{equation}
with a common transition baryon chemical potential $\mu_B^{\rm tr}$. The pressure at this point can define the transition pressure~\cite{Wu17,Bai23},
\begin{equation}
P^{\rm tr}=P^H(\mu_B^{\rm tr})=P^Q(\mu_B^{\rm tr}).
\end{equation}
Under the Maxwell construction, although both the chemical potential and pressure remain continuous, the number and energy densities of the two phases are not continuous at the phase transition point $(\mu_B^{\rm tr},P^{\rm tr})$, i.e.~\cite{Iva22},
\begin{equation}
\rho_B^H(\mu_B^{\rm tr})\neq \rho_B^Q(\mu_B^{\rm tr}), \qquad \varepsilon^H(\mu_B^{\rm tr})\neq \varepsilon^Q(\mu_B^{\rm tr}).
\end{equation}
This finite discontinuity is the hallmark of a first-order hadron-quark phase transition in the Maxwell construction~\cite{Sen24}.  It is worth mentioning that, in general, the $P(\mu_B)$ curves of the two phases may intersect multiple times. For the case of a single intersection, the hadronic phase is more stable at low densities, where $\mu_B^H < \mu_B^Q$ at $P = 0$. This implies that the energy per nucleon at zero pressure satisfies $\varepsilon^H/\rho_B^H< \varepsilon^Q/\rho_B^Q$, which is consistent with the neutron star or hybrid star scenario. Conversely, if $\mu_B^H > \mu_B^Q$ at $P = 0$, the quark phase is more stable at low densities, which supports the strange quark star hypothesis~\cite{Bod71,Wit84,Hol18}. As can be seen from Fig.~\ref{fig1}, even without the vector interaction ($G_V=0$), the baryon chemical potential of the quark phase at zero pressure is larger than that of the hadronic phase for different axion fields. This is suitable for constructing the hybrid-star equation of state, where the hadronic EOS is used below the transition density $\rho_B^H(\mu_B^{\rm tr})$, the quark EOS is used above $\rho_B^H(\mu_B^{\rm tr})$, and the transition part is replaced by a constant-pressure segment at $P^{\rm tr}$.

Finally, the crust of hybrid stars, in our calculations, is considered to be divided into two parts: the inner and the outer crust as in the previous treatment~\cite{Xu09a,Xu09b}. In the inner crust, a parametrized EOS of $P = a + b \varepsilon^{4/3}$ is used, and the outer crust, which usually consists of heavy nuclei and an electron gas, is described by the EOS from Ref.~\cite{Bay71}. Once the full hybrid-star equation of state is obtained, we can solve the Tolman-Oppenheimer-Volkoff (TOV) equations~\cite{Tol39,Opp39} for spherically symmetric, static stars to derive the mass-radius relation of the hybrid star. For the quark core, when the central pressure exceeds the transition pressure $P^{\rm tr}$, quark matter begins to appear at the center of the compact star. As the central density further increases, the quark core expands outward, and the resulting density jump across the two phases directly affects the mass-radius relation and the internal composition of the hybrid star~\cite{Sen24}.

\begin{figure*}[tbp]
\centering
\includegraphics[width=0.9\textwidth]{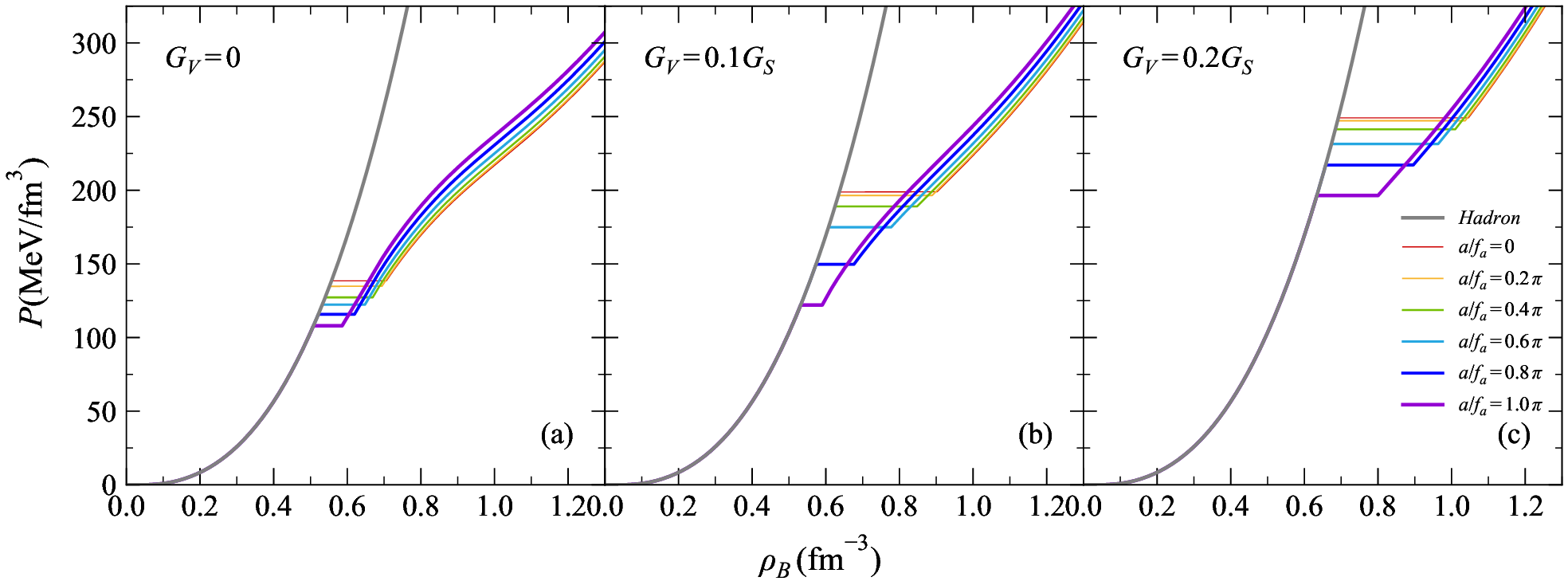}
\caption{Equation of States of hybrid star matter within the Maxwell construction. The grey solid line represents the pure, and the colored solid lines represent the quark phase with different scaled axion field $a/f_a$. Panels (a), (b), and (c) correspond to different vector couplings $G_V = 0$, $0.1G_S$, and $0.2G_S$, respectively.}
\label{fig2}
\end{figure*}

\section{Results and Discussion}
This work mainly focuses on the effects of the axion field on quark matter and the hadron-quark phase transition in hybrid stars. Therefore, the hadronic EOS based on the ImMDI model is held fixed. As shown in Fig.~\ref{fig2}, we present the EOS of hybrid star matter within the Maxwell construction, where the grey solid line (fixed) represents the hadronic phase, and the colored solid lines represent the quark phase with different values of the scaled axion field $a/f_a$. Panels (a), (b), and (c) correspond to different vector couplings $G_V = 0$, $0.1G_S$, and $0.2G_S$, respectively. In the NJL model, the scaled axion field $a/f_a$ serves as the chiral rotation angle of the KMT term and varies periodically~\cite{Cha12,Lu19,Liu25}. Therefore, the range of $a/f_a$ is taken as 0 to $\pi$ in this work.  As shown in Fig.~\ref{fig2}(a), the EOS of the high-density quark phase in hybrid stars becomes slightly stiffer with increasing $a/f_a$. This behavior can be understood from Fig.~\ref{fig1}. A larger $a/f_a$ causes quark matter to appear earlier at low baryon chemical potentials at zero pressure, thus stiffening the EOS of quark phase. Therefore, the intersection point of the two phases in $P(\mu_B)$ occurs at a lower baryon chemical potential and pressure under the effect of the axion. Accordingly, we can also clearly see in Fig.~\ref{fig2}(a) that the axion field causes the plateau (onset) of the first-order phase transition to shift significantly to lower densities and pressures. Interestingly, although the axion field slightly stiffens the EOS of quark matter, the earlier appearance of the first-order phase transition at low densities actually softens the EOS during the transition. These two effects compete in shaping the EOS of hybrid star matter and also affect the maximum mass of hybrid stars. This is different from the effect of the vector interaction. Comparing with Figs.~\ref{fig2}(b) and (c), we see that the vector coupling constant $G_V$ significantly stiffens the quark EOS, and the EOS becomes even stiffer with increasing density, which can be understood from the $\sum_i G_V\rho_i^2$ term in the thermodynamic potential of Eq.~(8). In contrast to the axion coupling, the vector interaction not only stiffens the quark EOS but also shifts the plateau (onset) of the first-order hadron-quark phase transition to higher density and pressure, thus making the entire EOS of hybrid star matter stiffer.

\begin{figure*}[tbh]
\centering
\includegraphics[width=0.9\textwidth]{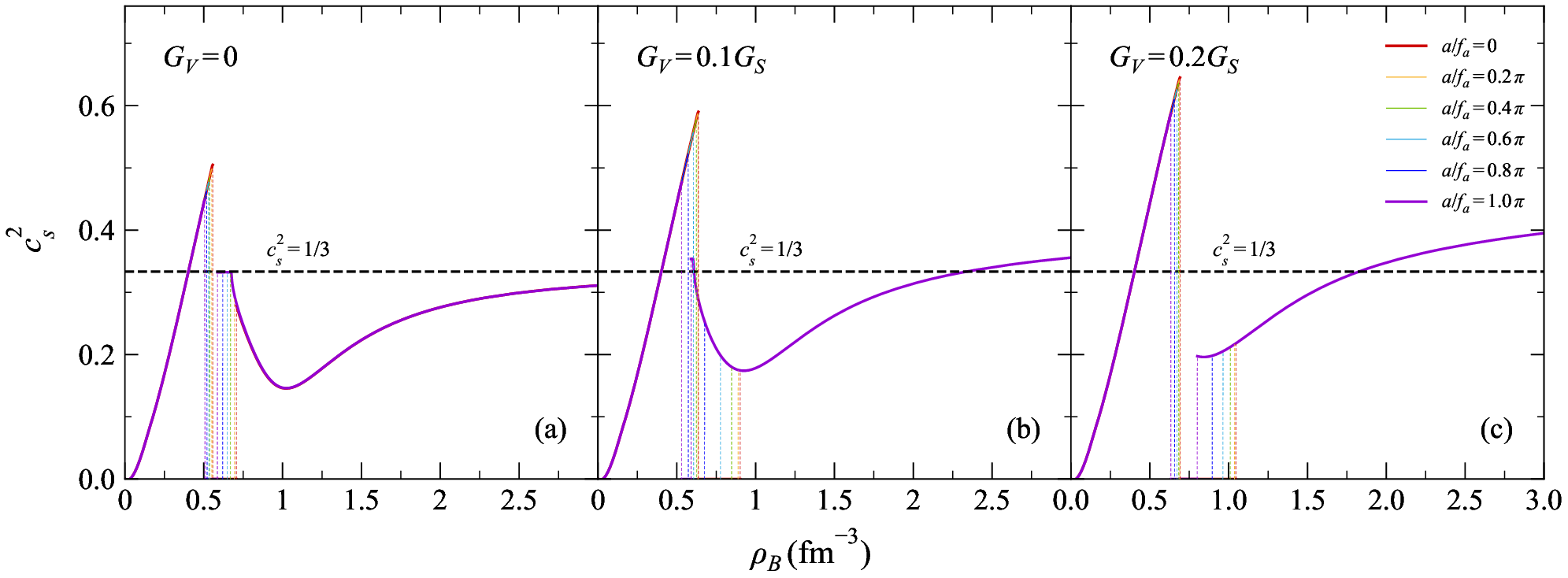}
\caption{Squared speed of sound $c_s^2$ as a function of baryon density for different values of $a/f_a$ and $G_V$. The dash line $c_s^2=1/3$ indicated the high-density conformal matter limit is shown for comparison.}
\label{fig3}
\end{figure*}

\begin{figure*}[htb]
\includegraphics[width=0.9\linewidth]{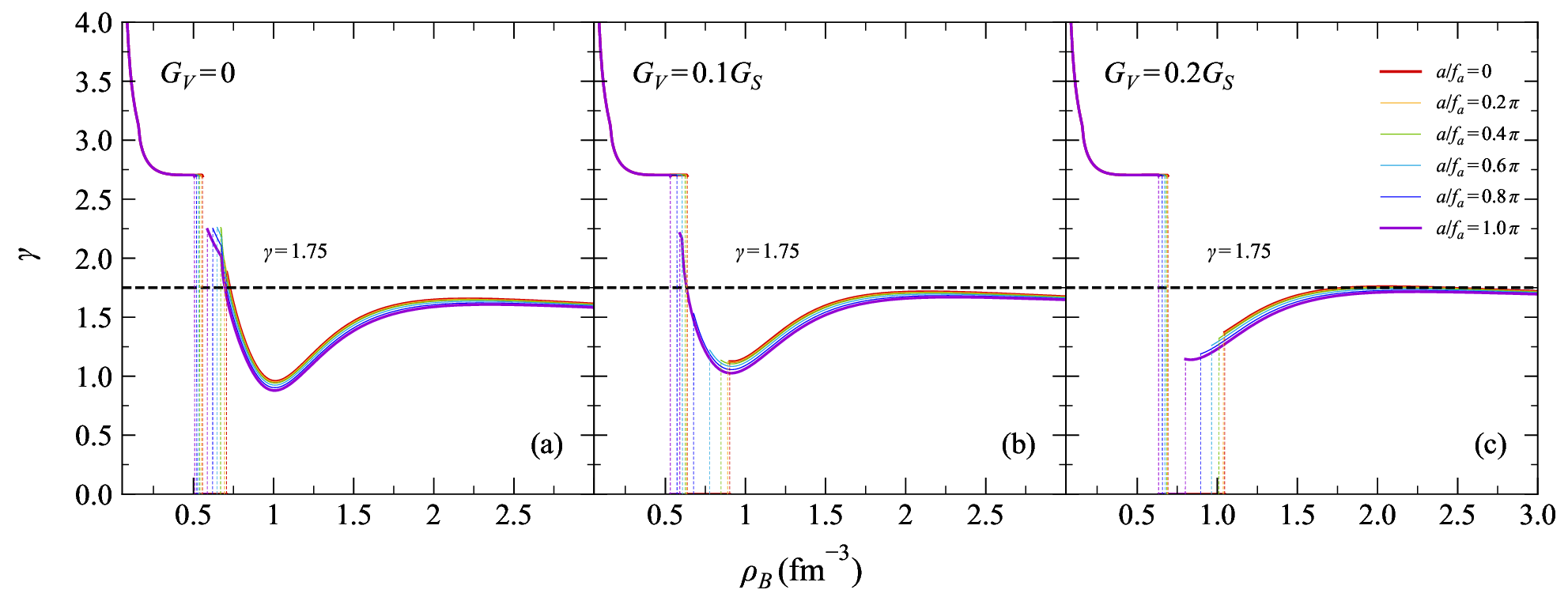}
\centering
\caption{Polytropic index $\gamma$ as a function of baryon density for different values of $a/f_a$ and $G_V$. The dashed line $\gamma \leq1.75$~\cite{Ann20} indicates the approximate criteria for distinguishing hadronic matter from quark matter.}
\label{fig4}
\centering
\end{figure*}
Using the obtained equation of state, we can determine other properties of hybrid star matter, such as the speed of sound and the polytropic index, which are respectively defined as $c_s^2 \equiv \partial P / \partial \varepsilon$ and  $\gamma \equiv d(\ln P) / d(\ln \varepsilon)$. Both $c_s^2$ and $\gamma$ can provide direct measures of the change in compressibility. The behavior of $c_s^2$ in hadronic matter is mainly characterized by the fact that below saturation density ($\rho_0=0.16$fm$^{-3}$), most hadronic models (such as chiral effective field theory) show that $c_s^2 \ll 1/3$, while at higher densities, the maximum value of $c_s^2$ is predicted to exceed 0.5~\cite{Gan09,Tew13}. Compared with hadronic matter, the speed of sound in quark matter is known to exhibit markedly different properties. For example, quark matter at very high densities ($\rho_B \sim 40\rho_0$) is approximately scale invariant or conformal, with the squared speed of sound $c_s^2 = 1/3$, whereas in hadronic matter the number of degrees of freedom is smaller and scale invariance is also broken by chiral symmetry breaking. Similarly, the polytropic index has the value $\gamma=1$ in conformal matter, while the hadronic models generically predict $\gamma \approx 2.5$ around the saturation density~\cite{Kur10}.  Therefore, $c_s^2$ and $\gamma$ are considered to be good approximate criteria for the evidence of quark matter in neutron stars~\cite{Ann20,Han23,Liu23,Yan24}. In Figs.~\ref{fig3} and~\ref{fig4}, we show respectively the squared speed of sound $c_s^2$ and the polytropic index $\gamma$ as functions of the baryon density in hybrid star matter by varying $a/f_a$ and $G_V$. It can be seen from the figures that at saturation density, the squared speed of sound satisfies $c_s^2 \ll 1/3$, the polytropic index $\gamma$ is close to 3, and the maximum values of $c_s^2$ in hadronic matter are also near or above 0.5, which is nearly consistent with the results of mainstream hadronic models. For the high-density quark phase, as shown in Figs.~\ref{fig3}(a) and~\ref{fig4}(a), the axion field has a very mild effect on the speed of sound and the polytropic index. Even as $a/f_a$ increases, $\gamma$ decreases slightly, indicating that the effect of the axion on the EOS (the increase in pressure) is almost independent of density or energy density. From the perspective of the speed of sound and the polytropic index, the primary role of the axion is to significantly shift the onset and the density range of the first-order hadron-quark phase transition. Of course, our results also indicate that if $G_V=0$, both $c_s^2$ and $\gamma$ could converge to the conformal limit at high density.  However, if $G_V \neq 0$, the strong repulsive effect from the vector interaction not only shifts the phase transition but also stiffens the EOS, driving $c_s^2$ and $\gamma$ to increase and even exceed the conformal limit of a free Fermi gas at high densities. Anyway, for all chosen parameter sets, our results are close to the approximate rules that $\gamma \leq 1.75$~\cite{Ann20} or $\gamma \leq1.6$ and $c_s^2 \leq 0.7$~\cite{Han23} can be used as criteria for distinguishing hadronic matter from quark matter. 

\begin{figure*}[htbp]
\includegraphics[width=0.9\textwidth]{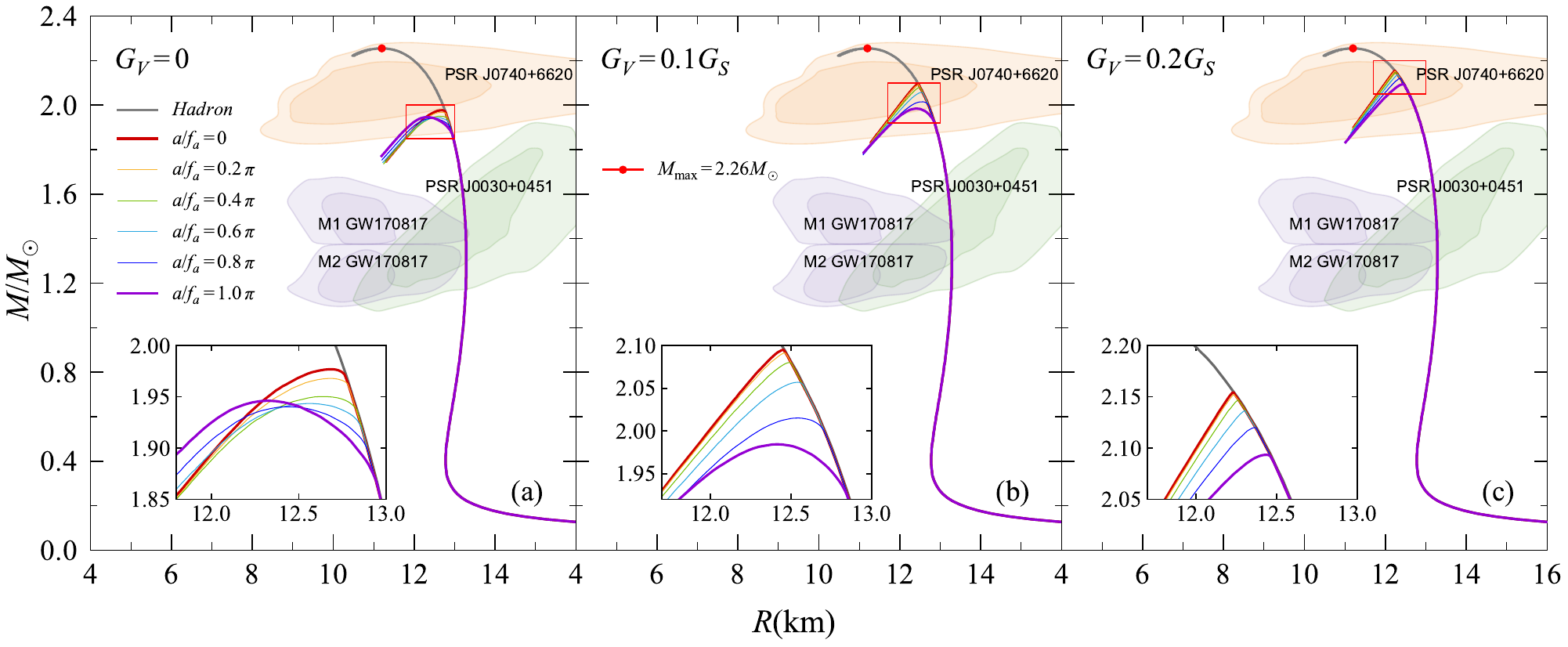}
\centering
\caption{Mass-radius relations of hybrid stars for different values of the scaled axion field $a/f_a$ and vector coupling $G_V$. The grey solid line represents the neutron star composed of pure hadronic matter and the red dot marks its maximum mass. The constraints from the Bayesian analyses of the observational data from the pulsars PSR J0030 + 0451~\cite{Ril19,Mil19} and PSR J0740 + 6620~\cite{Cro20,Ril21,Mil21}, and from the analyses of the gravitational-wave signal from the neutron star merger GW170817~\cite{Abb17} are shown for comparison.}
\label{fig5}
\centering
\end{figure*}

The changes in the EOS induced by the axion field and the vector interaction can be more intuitively reflected in the structure of hybrid stars. We show in Fig.~\ref{fig5} the mass-radius relations of hybrid stars for different values of $a/f_a$ and $G_V$. In each panel, the grey curve denotes the pure hadronic star, whose maximum mass reaches $2.26 M_\odot$, while the colored curves represent the corresponding hybrid star with hadron-quark phase transition. For comparison, the shaded regions show the constraints derived from Bayesian analyses of the pulsars PSR J0030+0451~\cite{Ril19,Mil19}  and PSR J0740+6620~\cite{Cro20,Ril21,Mil21}, and from the gravitational wave signal of the neutron-star merger GW170817~\cite{Abb17}. One can see that masses and radii of hybrid stars in all parameter sets are mostly consistent with these multi-messenger observational constraints. The results in Fig.~\ref{fig5}(a) show that the observed maximum mass of hybrid stars and the corresponding radius do not exhibit a clear monotonic dependence on $a/f_a$. This is due to the maximum mass of hybrid stars primarily constraining the EOS near the hadron-quark phase transition at densities of $2-5\rho_0$~\cite{Liu22}. As discussed above, in this density region, the axion field slightly stiffens the EOS of quark phase but simultaneously lowers the transition pressure. The competition between these two effects is also reflected in the maximum mass and the corresponding radius of the hybrid star.  For $G_V \neq 0$ (Figs. 5(b) and (c)), the vector interaction pushes the onset of quark appearance to higher densities, even beyond $5\rho_0$, which reduces the contribution of the quark-phase EOS to the maximum mass. Consequently, the softening effect of the axion field on the EOS of the first-order phase transition dominates and reduces the maximum mass of hybrid stars.
\begin{figure*}[htbp]
\centering
\includegraphics[width=0.8\textwidth]{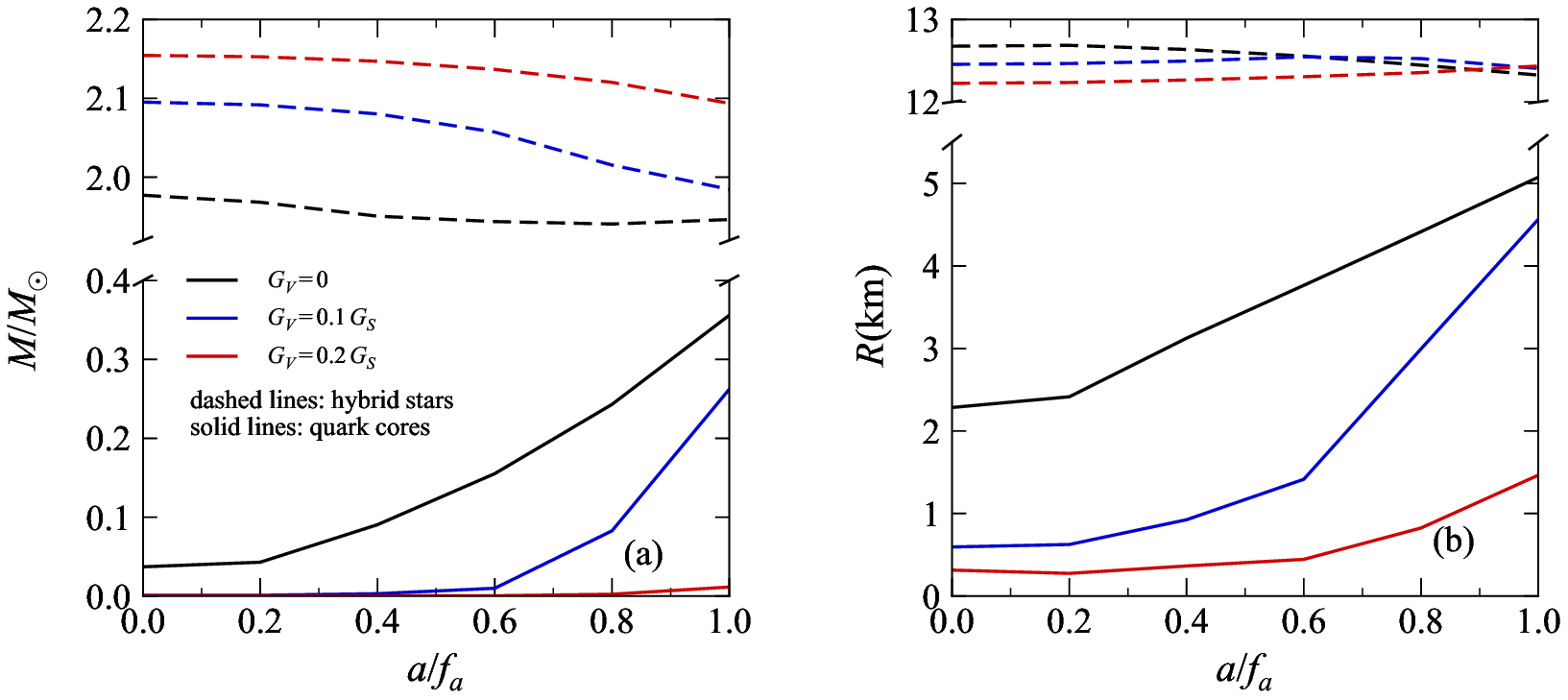}
\caption{Masses and radii of maximum-mass hybrid stars as well as the corresponding masses and radii of their quark cores as functions of the scaled axion field $a/f_a$ for different values of $G_V$. The dashed lines represent the maximum-mass hybrid stars, and the solid lines represent the corresponding quark cores.}
\label{fig6}
\end{figure*}

To better understand the axion effects on the structure of hybrid stars, we plot in Fig.~\ref{fig6} the masses and radii of maximum-mass hybrid stars as well as the corresponding masses and radii of their quark cores as functions of the scaled axion field $a/f_a$ for different values of $G_V$. The quark cores can be determined using the TOV equations by integrating them from the central baryon density to the onset of the quark phase.  As can be seen from the figure, for different parameter sets, the maximum masses of hybrid stars are all close to or exceed $2M_\odot$, and their corresponding radii lie in the range of 12-13 km. The dominant influence on the maximum mass comes from the vector interaction. As $G_V$ increases, the maximum mass increases significantly. In contrast, the axion field has only a very weak effect, especially at $G_V = 0$, where only a very minor fluctuation is observed. The radius of the maximum-mass hybrid star is mainly affected by the onset density of the quark phase. When $a/f_a = 0$ is fixed, increasing $G_V$ raises the onset density of the quark phase, leading to a decrease in radius. Conversely, when $G_V = 0$ is fixed, increasing the axion field lowers the onset density, leading to an increase in radius. When both  $a/f_a$ and $G_V$ are present simultaneously, no clear trend in the radius is observed.  Importantly, although the axion has a minimal impact on the mass and radius of the maximum-mass hybrid star, even simultaneously reducing both quantities slightly as $a/f_a$ increases (e.g., for $G_V=0$), it significantly increases the radius and mass of the quark-matter core. This is due to the axion effect substantially lowering the onset density of the quark phase and allowing the quark core to earlier appear at lower densities. Especially when $G_V = 0$, the mass of the quark core increases from $0.04M_\odot$ at $a/f_a = 0$ to $0.36M_\odot$ at $a/f_a = \pi$, while the radius of the quark core grows from 2.3 km to 5.1 km, approaching half of the total radius of the star. These results indicate that, within the Maxwell construction, the primary role of the axion is to promote an earlier hadron-quark transition in hybrid-star matter and to enlarge the quark core. Therefore, as a result of axion effects, a sizable quark core could appear in $2M_\odot$ massive neutron stars. Meanwhile, we also note that as $G_V$ increases, the quark core shrinks significantly and may even disappear. When the quark core vanishes, the effect of the axion on the hybrid star also disappears accordingly.

As a complement, the LIGO/Virgo Collaboration, in the analysis of the binary neutron-star merger event GW170817, provided a tidal deformability constraint for a $1.4M_\odot$ neutron star of $70 \le \Lambda_{1.4} \le 580$ at 90\% confidence level. However, it can be clearly seen from Fig.~\ref{fig5} that for all the parameter sets considered in our calculations, the $1.4M_\odot$ stars do not contain a quark core. Therefore, the axion has no effect on $1.4M_\odot$ neutron stars. In our previous work~\cite{Liu22}, we have already discussed in detail the influence of the parameters $x$, $y$, and $z$ on the tidal deformability of $1.4M_\odot$ stars. Hence, we do not elaborate further on this point in the present paper.

\section{Summary and outlook} 
In this work, we systematically investigate the effects of the axion field on the hadron‑quark phase transition and the properties of hybrid stars within the Maxwell construction. The hadronic phase is described by the ImMDI model with a fixed parameter set, and the quark phase by the three‑flavor NJL model with an axion‑modified KMT interaction. Our results show that the scaled axion field $a/f_a$ slightly stiffens the EOS of quark matter, but more importantly, it shifts the hadron-quark phase transition to significantly lower baryon densities and pressures. These two effects compete in shaping the EOS of hybrid star matter.  In contrast to the axion coupling, the vector interaction not only stiffens the quark EOS but also shifts the onset of the first-order hadron-quark phase transition to higher density and pressure. As a result, the axion field has only a weak influence on the maximum mass of hybrid stars, but significantly increases the mass and radius of the quark core, especially for weak vector coupling.  For  the case with $f_a=\pi$ and $G_V=0$, the mass of the quark core can reach up to 0.36 $M_\odot$, and the radius of the core exceeds 5 km, approaching half of the total radius of the star.  In the study of the speed of sound and polytropic index in hybrid star matter, we also find that the primary effect of the axion is to induce the hadron-quark phase transition to occur earlier at lower densities, while it has almost no effect on the speed of sound in the quark phase, even slightly reducing the polytropic index $\gamma$. This implies that the stiffening of the EOS (or the increase in pressure) due to the axion field is density-independent. Therefore, these results indicate that, within the Maxwell construction, the primary role of the axion is to promote an earlier hadron-quark transition in hybrid star matter and to enlarge the quark core. 

In the introduction, we also mentioned that the hadron-quark phase transition in this work is described based on the Maxwell construction. In studies of hybrid stars, the Gibbs construction is also widely used. However, both the Maxwell and Gibbs constructions only involve bulk contributions, neglecting finite-size effects such as surface and Coulomb contributions. Therefore, it is of interest in future work to explore the effect of the axion field on the hadron-quark phase transition including finite-size effects. Considering axions as cold dark-matter candidates, our results complement research into the impacts of beyond-Standard-Model particles on NS structure and binary mergers. As multi-messenger observations continue to improve, more precise constraints on the tidal deformability and radii of neutron stars will become available. Combining these data with theoretical predictions may allow us to determine the range of the axion-quark coupling strength, and even discover other new physics signatures beyond the Standard Model in dense matter.

\begin{acknowledgments}
This work was supported by the National Natural Science Foundation of China (Grant Nos. 12205158, 12575134, 12305148 and 12505159), the Shandong Provincial Natural Science Foundation, China (Grant Nos. ZR2021QA037, ZR2022JQ04, and ZR2025QC1487), and the Qingdao Natural Science Foundation (Grant No. 25-1-1-4-zyyd-jch).
\end{acknowledgments}

\end{document}